**Associative Memory For Reversible Programming and Charge Recovery**
Professor John Robert Burger
California State University Northridge
February 2006

*Abstract –* Presented below is an interesting type of associative memory called toggle memory based on the concept of T flip flops, as opposed to D flip flops. Toggle memory supports both reversible programming and charge recovery. Circuits designed using the principles delineated below permit matchlines to charge and discharge with near zero energy dissipation. The resulting lethargy is compensated by the massive parallelism of associative memory. Simulation indicates over 33x reduction in energy dissipation using a sinusoidal power supply at 2 MHz, assuming realistic 50 nm MOSFET models.

**Introduction**
One of the methods proposed to reduce average power dissipation within very large-scale integrated circuits is to recover the charge that is sent into the circuit. However, circuits must be structured to allow electrons to leave the circuit as readily as they enter. Also conceivable is a power supply that charges nearly as readily as it discharges. Systems whose goal is to operate using zero energy or zero average power are termed adiabatic, although near adiabatic is more accurate. Practical adiabatic circuits are discussed below with reference to a new type of memory word.

**Background**
Although publications on adiabatic circuits are sparse, the subject is important because of the excessive heat generated in VLSI (very large scale integrations) [1]. As a mundane example of charge recovery, consider a high Q resonant circuit. Once charged, it oscillates on its own, periodically returning charge to the capacitor. The objective below is digital circuitry that oscillates on its own, so that little or no energy is left at the location the digital circuit. The quality of the power supply is beyond the scope of this paper, except to say that with appropriate resonant properties, heat at its location may also be greatly reduced.

Consider a simple model of a CMOS gate as in Figure 1. Resistor R represents resistance in the channel of a MOSFET; capacitor C represents load.

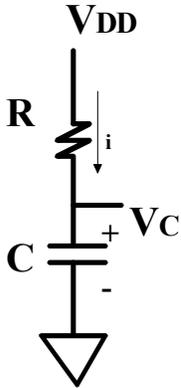

**Figure 1. Simple Model of a CMOS System**

A logic event that results in the charging of C can be modeled by letting $V_{DD}$ be a step function, that is, $V_1 u(t)$. During charging, equal amounts of energy enter into the capacitor and resistor (not proved here). Energy taken from the DC power supply is $CV_1^2$ while Capacitor energy is ½ $CV_1^2$. Today all of this energy is eventually converted into heat.

To accomplish charge recovery, note that resistor power dissipation is $i^2R$. If current i can be made small, then power lost in the resistance is insignificant. Let $V_{DD}$ to be a ramp that goes from 0 up to $V_1$ with a rise time of T as in Figure 2.

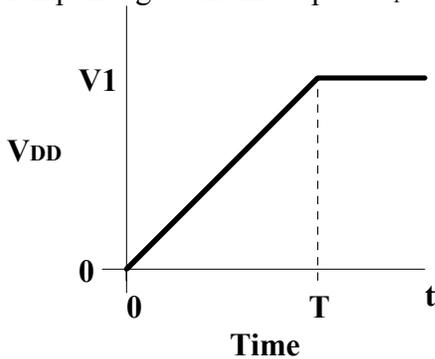

**Figure 2. Model of a Power Supply That Turns On Slowly**

Inaccurate, but useful is to assume that charging is so slow that capacitor voltage is about equal to the power supply voltage, that is, $v_C \approx V_{DD}(t)$. Then:

$$i \approx C \frac{dv_C}{dt} = C \frac{V_1}{T} \qquad (1)$$

Power lost in the resistor is:

$$P = i^2 R = \frac{RC}{T^2} C V_1^2 \qquad (2)$$

Energy lost in the resistor is:

$$E_R = \int_0^T P dt = \frac{RC}{T} C V_1^2 \qquad (3)$$

It is interesting to note that power dissipated in the resistor rapidly decreases for larger T. Indeed, all the energy taken from the power supply is given to the capacitor:

$$E = i \int_0^T V_{DD}(t) dt = \left( C \frac{V_1}{T} \right) \int_0^T \frac{V_1}{T} t \; dt = \frac{1}{2} C V_1^2 \qquad (4)$$

This model shows that charge recovery is possible if the rates of charging are controlled so that resistive losses are negligible. Once a capacitor is charged without significant charge dissipation, applying a negative edge, that is, V1 to zero with a fall time of T may similarly discharge it (therefore recharging the power supply). These equations neglect stray resistance to ground.

Note that power reduction in R is unrelated to the standard equation for CMOS power dissipation due to unregulated servicing of capacitance:

$$P_d \approx f_1 C V_{DD}^2 \qquad (5)$$

This equation supposes clocking with square waves of frequency $f_1$, so that all power goes to channel resistance. Under charge recovery, equation (5) is irrelevant. Ideally, Equation (2) applies. Power in Equation (2) is very much lower, and is proportional to the square of frequency assuming f = 1/2T.

Note that the energy in a capacitor is ½ $C V_1^2$ in Equation (4) not just for a ramp but also for any waveform between 0 and $V_1$. In particular, $V_{DD}(t)$ may be sinusoidal, that is, ½$V_1$ [1-cos(πt/T)], as assumed below.

**Circuit Design Principles**
Digital circuits can use a sinusoidal power supply; results are available at the peak of the voltage. Equal amounts of energy are supposed to be given to, and taken from the circuit. To accomplish this, it is required to:

1) Avoid turning on a transistor that has voltage across it
2) Avoid turning off a transistor that has current flowing through it.
3) Avoid creating trapped charge.

Trapped charge is charge that cannot be returned to a power supply, and represents a loss of energy (at least until it is returned). For example, the charge involved in a traditional sample and hold operation is trapped and will absorb energy. However, if charge can be accumulated slowly, a subsequent power supply cycle can be employed to retrieve it. Note that CMOS provides capacitive loads. A resistive load, or any stray resistance to ground, even if external, heats the integrated circuit.

One penalty for adiabatic logic is a slower system whose power supply frequency is f = 1/(2T). So although adiabatic integrated circuits can dissipate less power at the location of the circuit, they also operate slower. For digital circuits, parallel processing can overcome this problem. Assume, for example, that T is X times the RC time constant modeled in Equation (3). If Y processors operate in parallel, but they are X times slower, they have the potential to outperform a single processor by a factor of Y/X. Typically it is possible to have Y/X >> 1.

Without going into the thermodynamics involved, it has been known for years that a reversible computer has the potential to perform computations using zero energy and zero average power [2]. Toggle memory, described next, is capable of being a reversible computer.

**Toggle Memory**
The cells of interest can be defined with reference to Figure 3. It models a single word in a memory system possibly with billions of identical words, all of which are going to operate in parallel. Identical control lines, not shown, are assumed to string through each cell.

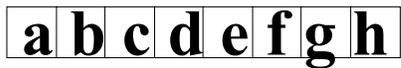

**Figure 3. Representation of a Word in Associative Memory**

This word can test a set of its own bits and if the tested bits are all true, then the word will complement bits within itself. For example, a command might be, if a, b, c, d, e, f and g are true, then complement h, where h is a flag that indicates that a keyword has been found. Note the revolutionary concept that bits are toggled (as opposed to being read and written). Ordinary random and associative memory require multi reads and multi writes for each bit, because each must be read, analyzed, possibly complemented, and written back. Conventional reading and writing, say for ½ billion words, is a major bottleneck. So there is motivation to design toggle memory for use in words that do what Figure 3 represents.

The above concept of toggling is the basis of a powerful machine whose programming is unimportant to this paper [3]. Logically reversible gates are well known [4]. However, any practical computer must also be massively parallel, as it would be with toggle memory, to overcome the drawback that electrically reversible gates are very slow. In toggle memory, irreversible instructions are possible, but they waste charge irreversibly. For example, a bit may test itself, and if true, it toggles itself to false. This type of operation is not logically reversible, because there is no history left to compute the bit's original value. Also it is not electrically reversible in toggle memory, so uses extra energy.

The task at hand is to demonstrate an appropriate toggle circuit, since one is not readily available, especially not one that incorporates charge recovery principles.  What follows is basic CMOS circuitry for a special purpose toggle circuit.   After presenting the basic toggle circuit, the most dissipative part of the circuit will be engineered for charge recovery.  The engineering will employ the author's novel method based on single-level, as opposed to split-level logic [5, 6].  The term split-level refers to trapezoidal pulsating power supplies that are centered so that one goes up to true; the other goes down to false. Circuits in this paper were simulated using shareware known as WINSPICE; the transistor models are the public-domain 50 nm BISIM4 models [7].

**Conditional Toggle Circuit Design and Simulation**
There are many ways to design a conditional toggle circuit, but the following approach is tailored for charge recovery from an inter word bus, that is, a bus that goes from cell to cell within the word.  This bus is also known as the matchline.  Figure 4 shows a simulation model.  The inter word bus is pre charged through transistor M1 located at the end of the word, and serving all cells in the word.  Next is the bit test: assume that the control signal $V_{FM}$ <u>pulses low</u>.  If the bit-value at the output of the cross-coupled flip flop M7, 8 is true, then the other output M9, 10 is false; the input to the 'inverter-AND' gate M3, 4 is false.  When $V_{FM}$ pulses low, transistor M2 remains off and the bus will not discharge.

On the other hand, if the output M9, 10 of the flip-flop is true, M2 conducts to discharge the bus irreversibly.  Simulations indicate that a $V_{DD} = 1$ V system discharges an assumed equivalent capacitance of 1 pF in less than 10 ns assuming L=100 nm and W=1 um in M2.

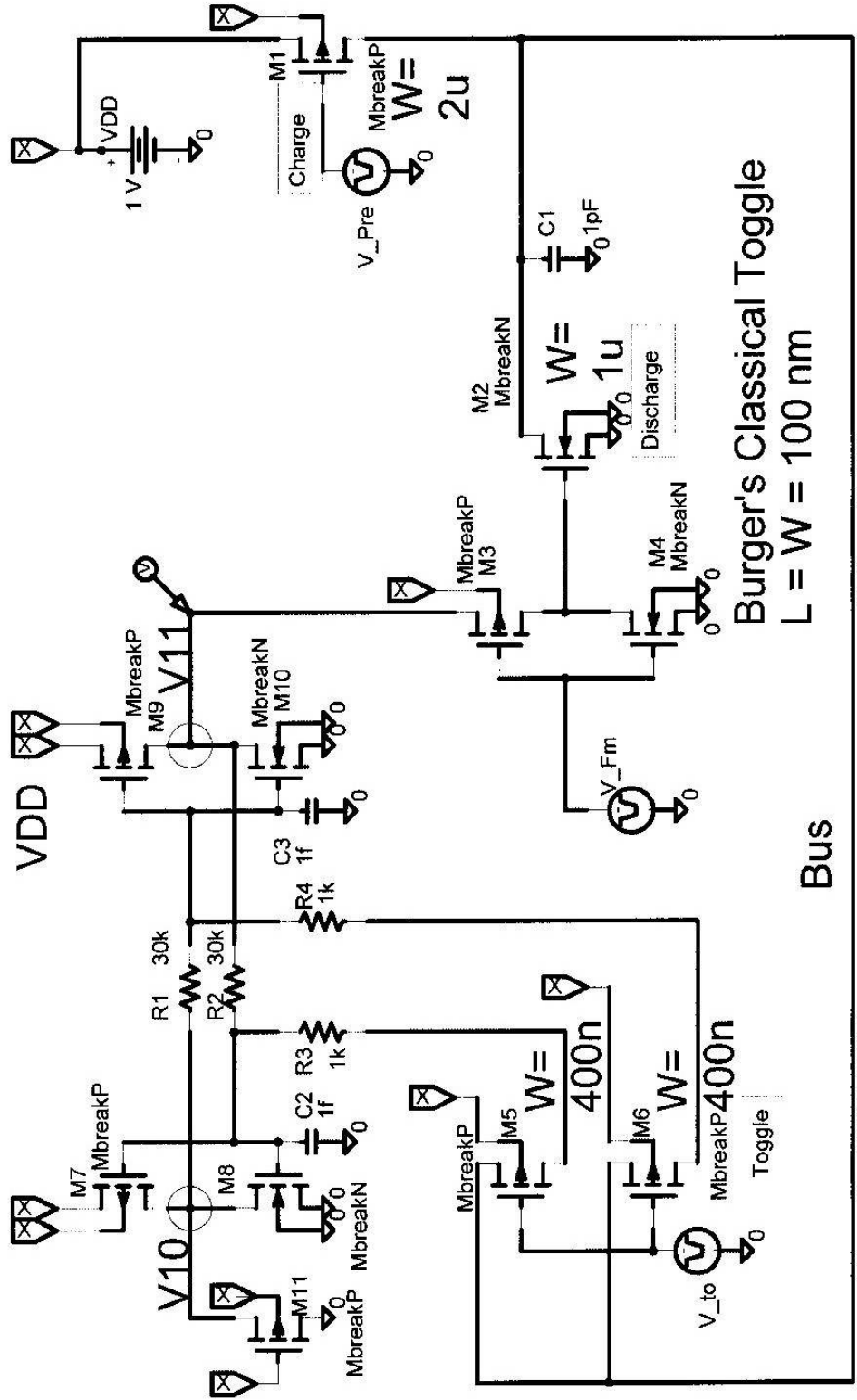

**Figure 4. Simulation Circuit of Conditional Toggle**

At this point $V_{TO}$ is high; triggering transistors (M5, 6) are cut off; the flip-flop is in a given state. Pulsing low the control signal $V_{TO}$ executes a toggle. This is a conditional toggle that works only if the bus is charged. $V_{TO}$ engages the P-switches (M5, 6) so that a sample of the bus voltage is placed on gates of the cross-coupled flip flop (M7, 8, 9 and 10). After the $V_{TO}$ pulse releases, a toggle occurs if the bus is charged. It is critical that the toggle circuit be balanced for reliable toggling, which is why M11 is added to counterbalance the load imposed by M3.

During the toggle pulse, resistors R1 and R2 permit the gate voltage to increase. But if they are too low in resistance, they prevent the gates of M7 and M8 from going high enough to toggle. Also, resistors R3 and R4 must be high enough in resistance to provide sufficient isolation of the right and the left gates. Actual resistance values were found to be non critical. Empirically, via simulation, it was found that the product of resistance (for example R1 in k$\Omega$) and transistor width (for example W5 in um) must be more than 4 in this model:

$$R_1\, W_5 > 4 \tag{6}$$

The product was set to 12 to create design margin, and yet be low enough so that components can be conveniently integrated.

In this circuit, the duration of the toggle signal must be above roughly 2 ns according to the simulation. Width cannot be much beyond 20 ns, to avoid excessive discharging of the bus voltage. A feature of the trigger circuit is that it only triggers if bus voltage is above about 70% of VDD, that is, about 0.7 V in the simulation. This means that toggling works only for bus voltages above about 70% of VDD. Typical simulated toggling is shown in Figure 5. Note the discharge of the bus voltage V(2) during each trigger pulse. Note also the reliable triggering of from high to low, and low to high in this one-volt system. Once bus voltage has dropped to about 0.7 volt, additional toggles are impossible.

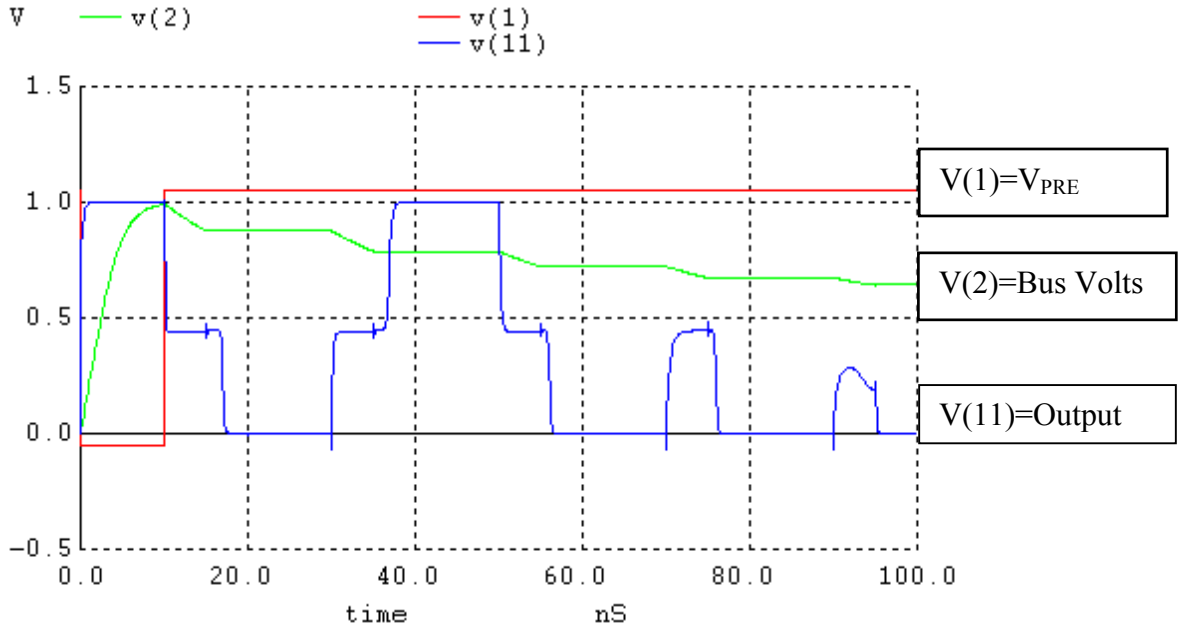

**Figure 5.   V(2) is bus voltage; V(11) is flip flop output**

To conclude this section, a properly designed trigger circuit does not require a full swing of logic from 0 to $V_{DD}$.  So the bus does not have to completely discharge to control the toggle, a fact that can result in significant savings in average power and energy.

**Charge Recovery Simulation**

It was decided to recover only the charge in the bus, that is, in the 1 pF capacitor, mainly because if readers are interested in anything, they will be interested in fixing what consumes the lion's share of energy.  The energy in the capacitor is 0.5 pJ, but the energy historically taken from $V_{DD}$ is double that, or 1 pJ, as can be verified by writing a simulator subroutine to numerically integrate the power from $V_{DD}$.  To accomplish charge recovery, a sinusoidal power supply is used for M1 only, as suggested in the simulation model shown in the upper right of Figure 6.  The voltage equation is:

$$V_{DD}(t) = 0.75 + 0.25 \sin [2\pi f (t + \delta)] \tag{7}$$

Frequency f=10 MHz, Half period T = 50 ns; and delay $\delta$=-75 ns to make the power supply voltage look like it goes from low to high, that is, a negated cosine waveform.  As a reasonable tradeoff, frequency was set to be 10x lower that the maximum clock rate of the bus.  Minimum rise and fall times were about 5 ns, so maximum clock rate is about 100 MHz.  All other circuits keep the same DC potentials.

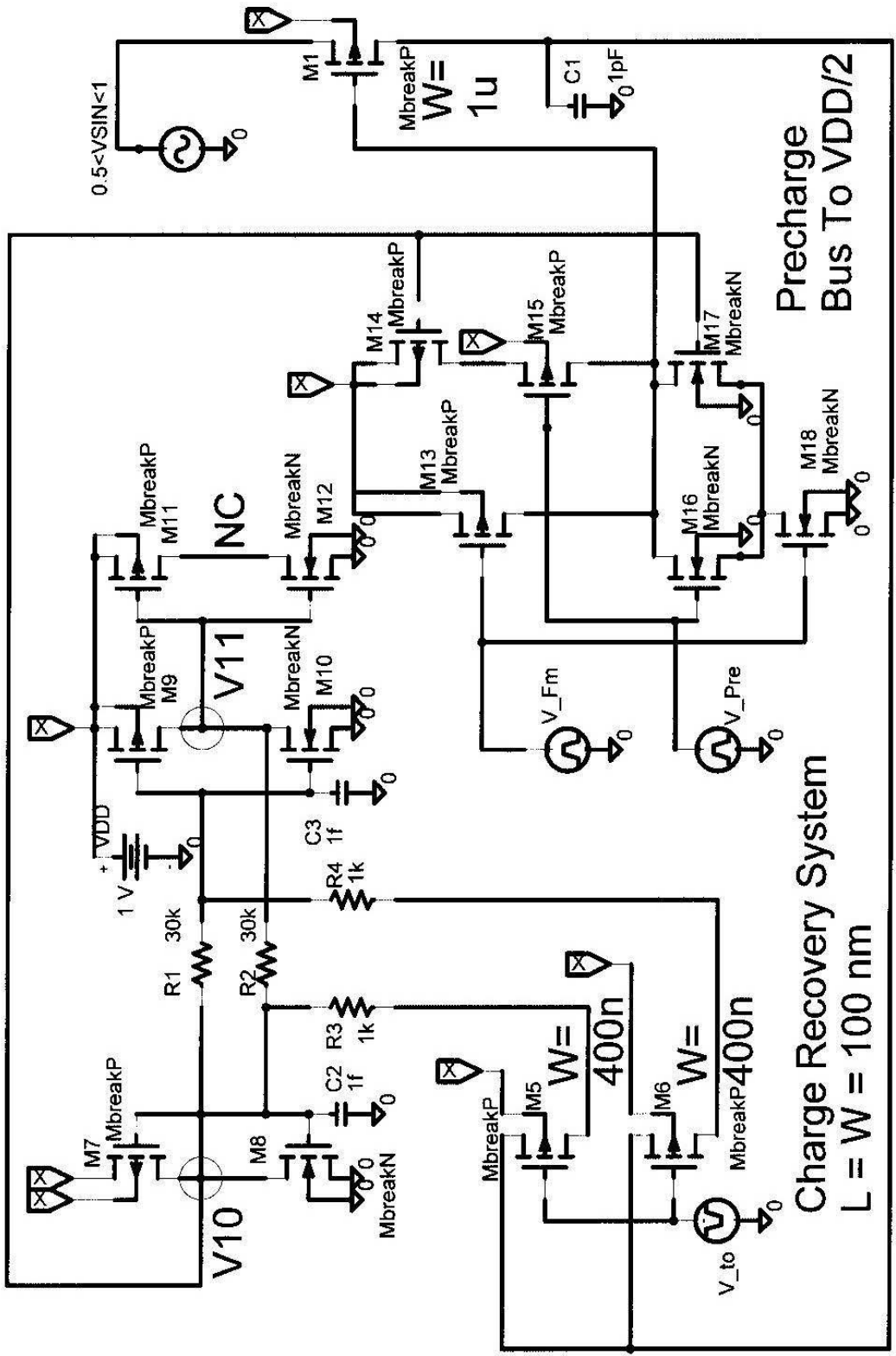

Figure 6. Simulation Circuit for Charge Recovery in C1

The charge recovery method can be explained with reference to the waveforms in Figure 7.

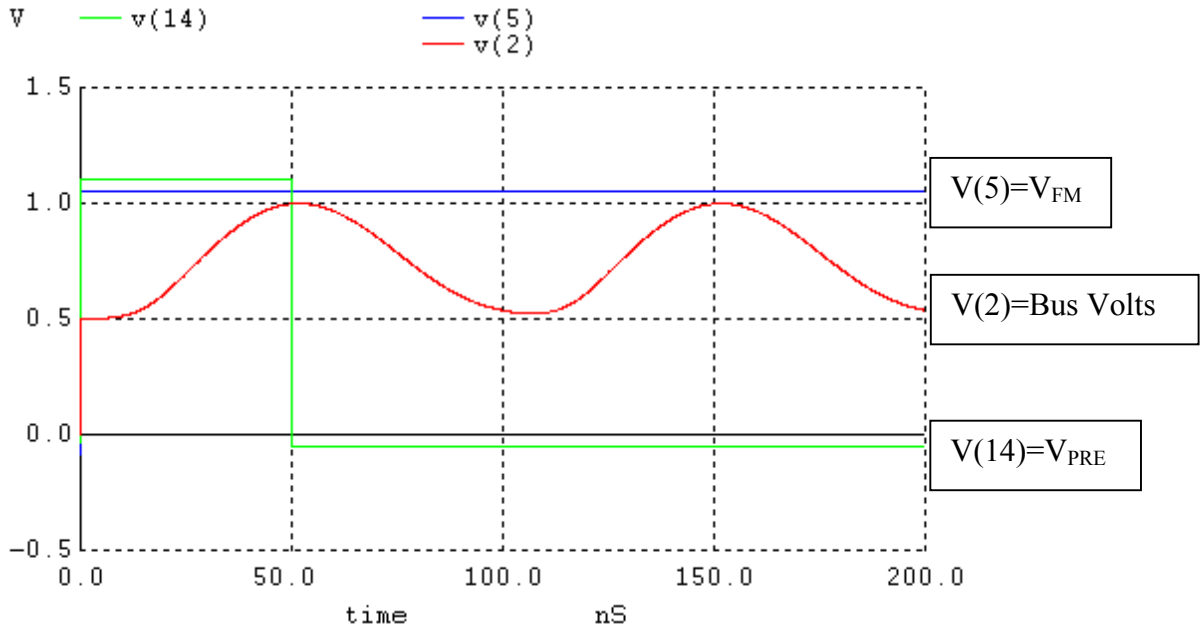

**Figure 7. Toggle output low; Bus charges first half cycle; discharges second half cycle**

The control signals $V_{FM}$, that is, v(5) and $V_{PRE}$, that is, v(14), explained later, permit bus V(2) to charge between 0 and 50 ns. When the toggle output V(11) at M9, 10 is low, corresponding to a tested bit being zero, the bus is discharged by the power supply, as indicated between 50 and 100 ns. In this design the bus voltage is assumed pre charged to ½ VDD, and does not need to go below this value. Bus voltage V(2) follows the sinusoidal power supply with a slight offset. The energy in femtojoules as given to, and taken from the bus system (transistor M1 width W1=2.5 um) is computed in Figure 8.

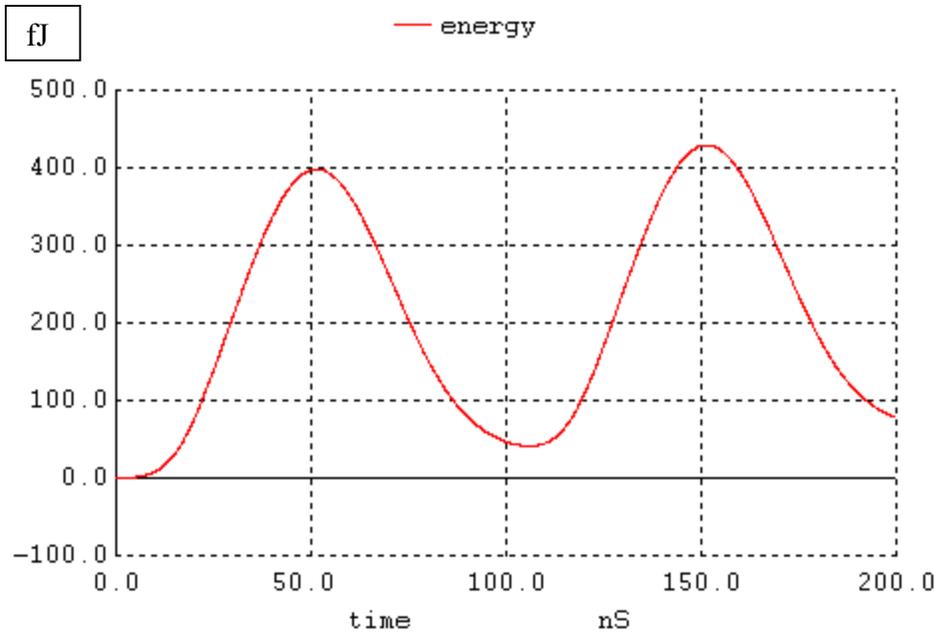

**Figure 8. Computed Energy as a Function of Time (Toggle Output $V_{11}$ False)**

Note first that at 50 ns the peak energy is about 397 fJ. This can be analyzed as an approximation to the energy increase in the capacitor: $\frac{1}{2} C(v_1^2 - v_0^2) = 0.5 - 0.125 = 375$ fJ. At 50 ns approximately 22 fJ is lost in the circuit at 10 MHz.

Between 50 and 100 ns, charge is allowed to return to the power supply. Therefore, the net energy dissipated by the circuit is only about 43 fJ at 100 ns. Two cycles are shown because two cycles are required to accomplish conditional toggling. At 200 ns the energy loss indicates about 77 fJ. Note that the energy values are not exact multiples. There are small errors because stray resistance that affects energy; also, for low energy values, the energy curve is slightly delayed relative to the power supply voltage, so the values given are not exact minimums; numerically, the integration uses a method of quadrature using the voltage and current vectors over time, so small numerical errors are expected. In this case, print step is 2 ns. Print step is held 100x below simulation time in all simulations below.

Figure 9 shows what happens when the toggle output is high.

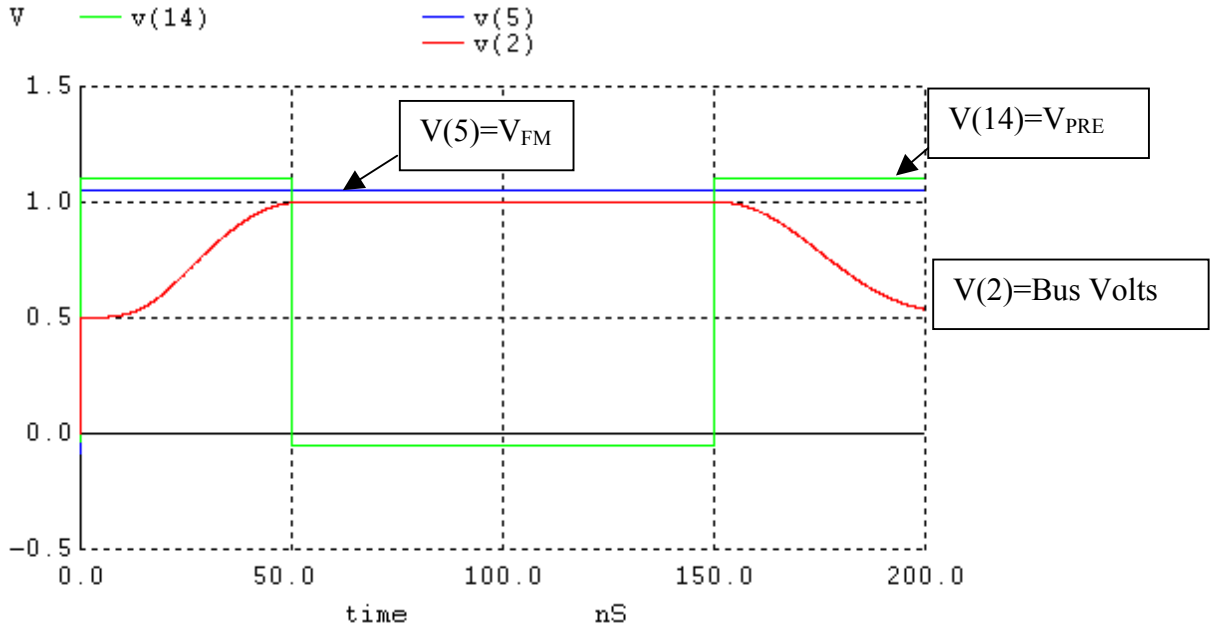

**Figure 9. Toggle output high; Bus charges first half cycle; discharges second half of second cycle**

When the toggle output V(11) at M9, 10 is high, the voltage between 50 and 150 ns is held.  In this region, near 100 ns, a trigger signal may occur.  It is important to note that the triggering signal is usually not sent to the same cells where  $V_{FM}$ signals are sent.  Otherwise there could be an irreversible bus discharge near 100 ns when the toggle output V(11) goes low.   Reversibly, the bus returns charge to the power supply on the second half of the second cycle, when $V_{PRE}$ returns high.  Figure 10 shows Energy consumption  over two cycles is 43 fJ.

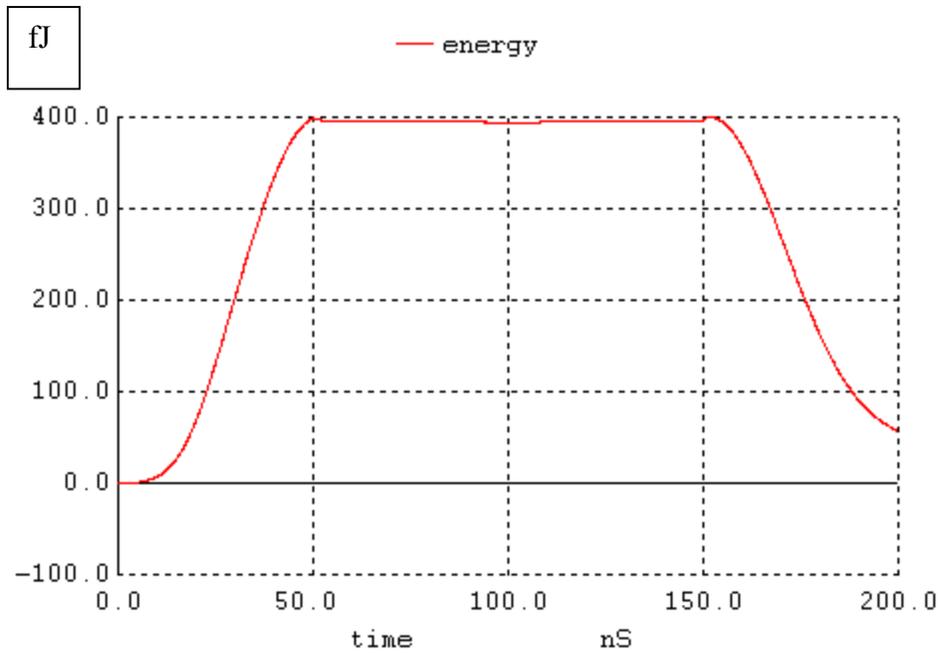

**Figure 10. Computed Energy as a Function of Time (Toggle Output $V_{11}$ True)**

If the toggle output was originally true, the circuit uses about half the energy (in two cycles) than when the output is false. As a note to interested readers, the physical width of transistor M1 affects energy dissipation. If this is made smaller, then energy for one cycle goes up as in Figure 11.

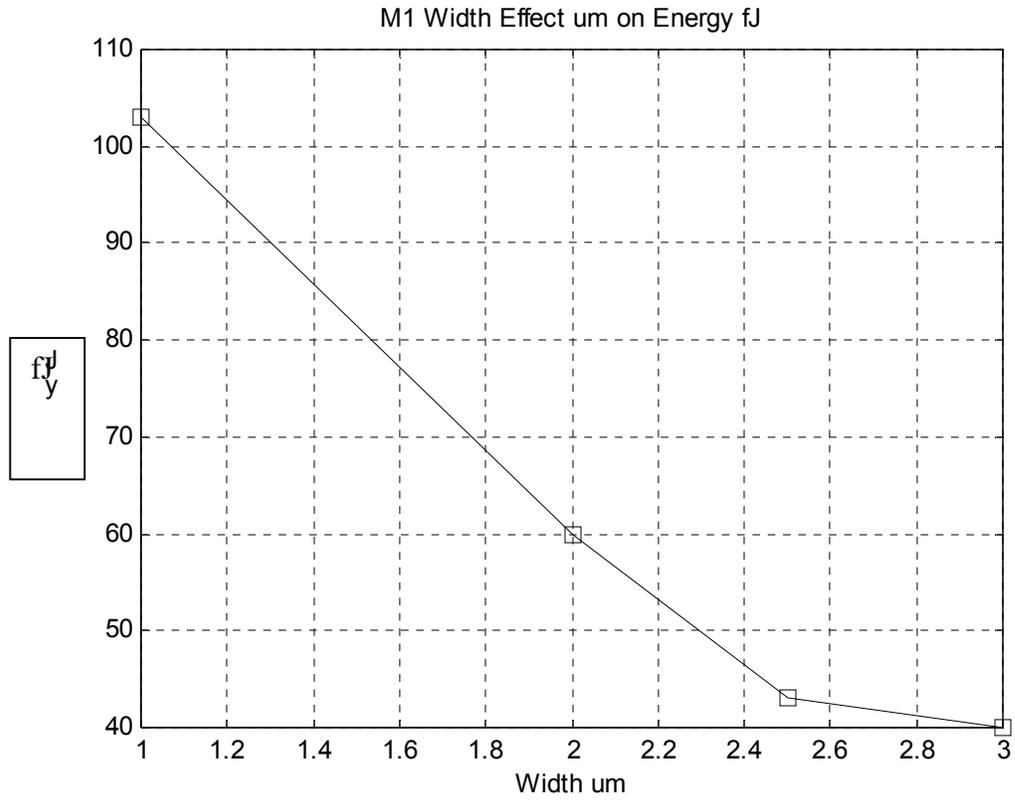

**Figure 11. M1 Width Effect**

The width of M1 was set to 2.5 um for low energy loss at 10 MHz.

**Charge Recovery Limitations**
Figure 12 shows simulated energy for two cycles as a function of power supply frequency:

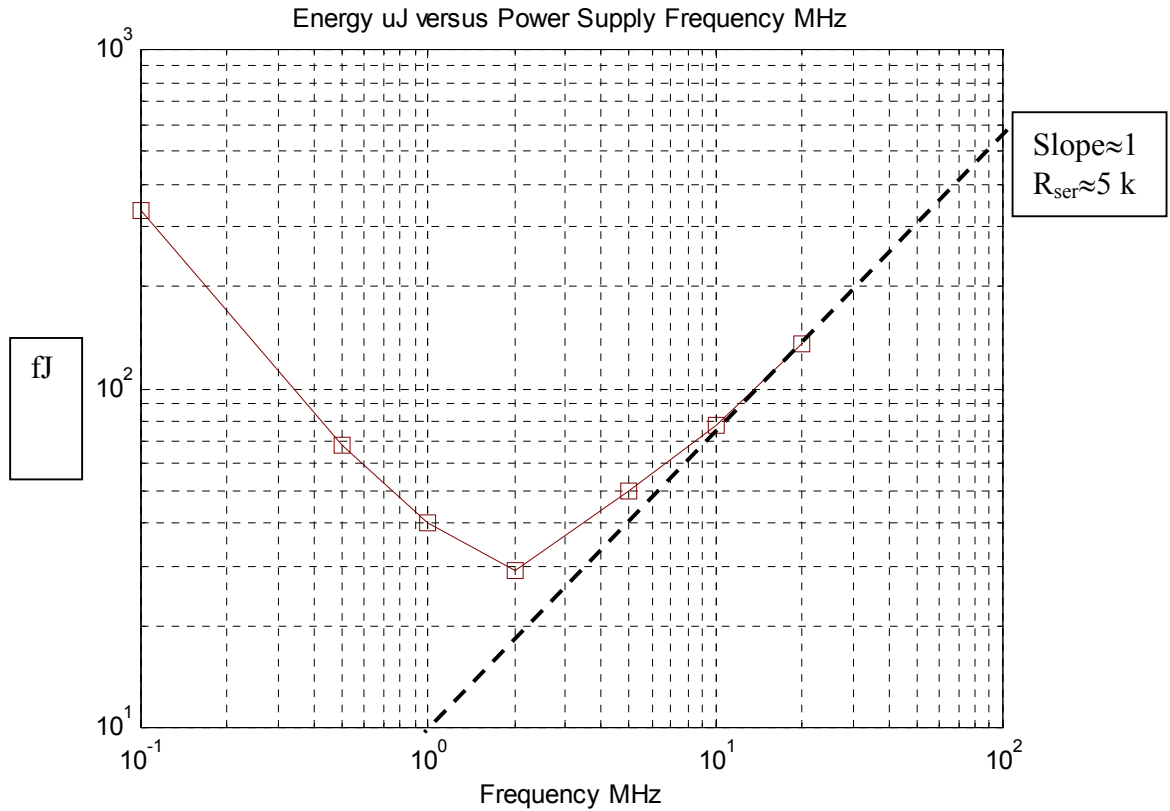

**Figure 12. Energy Dissipated (Two Cycles) Versus Power Supply Frequency f**

At higher frequencies, all energy is dissipated. An upper energy limit for two cycles would be 4 x 375 = 1500 fJ (not shown in Figure 12). At lower frequencies, it is conjectured that energy increases because of stray resistance in the circuit. This type of energy loss is expected to increase as 1/f. Loss to stray resistance is apparent in Figure 13, which shows the effect from an arbitrarily placed 1 MEG from the source of M1 to ground:

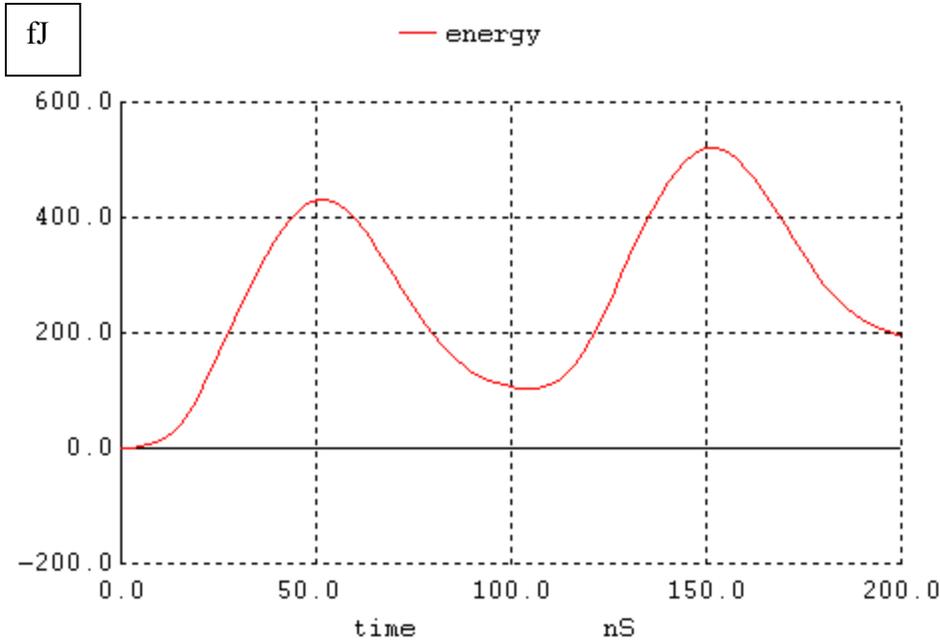

**Figure 13. Energy loss due to 1 MEG stray resistance at 10 MHz**

Note, compared to Figure 8, the curve is pulled upward with time. This is what one sees as power supply frequencies are lowered below 1 MHz in this model. Minimizing stray resistance, including sub threshold conduction, is important for charge control. However, much can be done immediately. Assuming the 50 nm models as given, energy can be made to drop by a factor of 33 at 2 MHz, which is significant.

**Application of the Simple Model**
The simple model represented by Equation (3) can be applied to the bus driver circuit. Note that f = 1/2T. The energy lost, expressed as an equation of logarithms, is:

$$Log_{10}E_R = Log_{10}f + Log_{10}\left(2R_{ser}C^2V_1^2\right) \qquad (8)$$

$E_R$ for two cycles can be in fJ, f can be in MHz, C can be in pF, and $R_{ser}$ can be in k$\Omega$. Assuming that energy lost in the resistor $R_{ser}$ is the energy actually dissipated, that is, neglecting stray resistance, the slope of a curve is the coefficient of $Log_{10}f$, that is, unity. The slope in Figure 12 between 10 and 20 MHz is about [$Log_{10}(136)$ - $Log_{10}(77)$] / [$log_{10}(20)$ - $Log_{10}(10)$] = 0.82. It is approaching unity.

For the purposes of predicting energy dissipation using Equation (8), assuming no stray resistance, series resistance $R_{ser}$ is 5 k. This is based on 10 fJ at 1 MHz in Figure 12, assuming C = 1 pF and $V_1$ = 1 V.

**Control Signals**
The control signals $V_{PRE}$, $V_{FM}$ and $V_{TO}$ are sent to all memory words in parallel. The logic for the gate of transistor M1, a P-switch with active low logic, must be:

$$G_{M1} = \overline{V_{FM}} + \overline{V_{PRE}} V_{11} \qquad (9)$$

If $V_{FM}$, that is, V(5) is true, and $V_{PRE}$, that is V(14) is true, the P-switch M1 closes to charge the bus for the first half cycle. Then $V_{PRE}$ goes false. If the toggle output $V_{11}$ = V(11) is false, the bus discharges during the second half cycle of the power supply. However, if V(11) is true, the bus holds its charge and is ready for a toggle signal from $V_{TO}$ to some other cell in a reversible system. The signal $V_{PRE}$ again goes high for the last half of the last cycle so that the bus can return its energy to the power supply.

If the signal $V_{FM}$ is false, M1 cannot conduct. An implementation of the logic of Equation (8) translates into the logic gate comprised of M13, 14, 15, 16, 17 and 18 in Figure 6.

**Conclusions**

Introduced above is toggle memory in which reversible instructions result in reduced energy usage. Thus the Feynman ideal is approached: a reversible computer with reversible programming that requires little or no energy to operate once started.

Any computer can be reversible, but it will not be adiabatic unless it contains reversible circuits. The bus circuit given above is an example of a reversible circuit in which resistive losses are driven toward zero, while capacitive energy is recovered. In fact, capacitive energy is available to recharge the power supply at its location. Most important is that unnecessary heat is not left at the location of the circuit.

A reversible circuit is accomplished by applying the above principles of adiabatic circuit design, including 1) never turning on a transistor that has voltage across it, 2) never turning off a transistor that has current flowing through it and 3) never permitting charge to become trapped. These principles are applied with the aid of a sinusoidal power supply directly driving the transistors.

The result is associative memory that supports both reversible programming and charge recovery. Simulated charge recovery resulted in over 33x lower energy usage by the matchline. What is appropriate about the toggle is that it requires only one sinusoidal power supply, one that moves bus voltage between ½ $V_{DD}$ to $V_{DD}$ to control triggering. This halves the energy dissipation, since split-level logic generally requires two pulsating power supplies.

Without significant parallelism, unfortunately, adiabatic logic would be impractical, since clocks run slower. But massively parallel circuits compensate. The use of toggle memory, described above, is a good application of charge recovery.

*Acknowledgement* – Many thanks to Professor Sembiam Rengarajan for helpful discussions about this article.